\documentclass{iau}

\usepackage{amsmath}
\usepackage{graphicx}
\usepackage{multirow}

\begin{document}

\lefttitle{D. Rodgers-Lee}
\righttitle{Stellar energetic particles and cosmic ray effects in exoplanetary atmospheres}

\jnlPage{1}{7}
\jnlDoiYr{2024}
\doival{10.1017/xxxxx}
\volno{388}
\pubYr{2024}
\journaltitle{Solar and Stellar Coronal Mass Ejections}

\aopheadtitle{Proceedings of the IAU Symposium}
\editors{N. Gopalswamy,  O. Malandraki, A. Vidotto \&  W. Manchester, eds.}

\title{Stellar energetic particle and cosmic ray effects in exoplanetary atmospheres}

\author{D. Rodgers-Lee}
\affiliation{Dublin Institute for Advanced Studies}

\begin{abstract}
Energetic particles, in the form of stellar energetic particles and cosmic rays, can lead to disequilibrium chemical effects in exoplanetary atmospheres. In Earth-like atmospheres, energetic particles can drive the formation of prebiotic molecules, the building blocks of life. Here instead, I study the transport of energetic particles through a hydrogen-dominated exoplanet atmosphere and calculate the resulting ionisation rate of molecular hydrogen using a Monte Carlo energetic particle transport model. I focus on a GJ436\,b-like atmosphere at orbital distances between 0.01-0.2\,au which includes the orbital distance of the exoplanet GJ436\,b (0.028\,au). I found that stellar energetic particles lead to high ionisation rates in a GJ436\,b-like atmosphere between 0.01-0.2\,au. These results motivate the use of chemical models of gas giant atmospheres including energetic particle ionisation to ultimately produce synthetic James Webb Space Telescope (JWST) and Ariel transmission spectra in the future.

\end{abstract}


\maketitle

\section{Introduction}
There are two types of energetic particles that are important for exoplanet atmospheres. These are stellar energetic particles from the coronal mass ejections and flares of low-mass stars like the Sun and Galactic cosmic rays that enter an astrosphere\footnote{An astrosphere is the stellar equivalent of the Sun's heliosphere.} from the interstellar medium (ISM). These two types of energetic particles then reach the location of an exoplanet. Here, they encounter any existing exoplanetary magnetosphere and interact with the atmosphere of the exoplanet. 

 Energetic particles are of interest in the context of exoplanet atmospheres because they can cause a number of interesting effects. They can drive the formation of prebiotic molecules in Earth-like atmospheres \citep{airapetian_2016, dong_2019} which are the molecules thought to be important for the origin of life on Earth. On the other hand, they can also cause fake biosignatures in Earth-like atmospheres \citep{grenfell_2012}. A biosignature is a chemical signature in an exoplanet atmosphere thought to be indicative of life. Thus, since energetic particles can also contribute to fake biosignatures it is important to constrain their effect so that it can be removed when searching for real biosignatures. Last, to isolate the chemical effect of energetic particles, \citet{helling_2019} and \citet{barth_2021} found that energetic particles can lead to the formation of exotic molecules in hydrogen-dominated atmospheres that are not expected to form otherwise which are known as fingerprint ions. 

 More broadly, studying the effect of energetic particles is important because high fluxes of energetic particles affect life-forms by damaging DNA \citep{herbst_2019,atri_2020}. It has even been suggested that energetic particles (in this case Galactic cosmic rays but the same logic applies to stellar energetic particles) indirectly left an imprint on the helicity of DNA \citep{globus_2020}.

 \section{Methodology}
 Here, I will focus on the impact of energetic particles in the GJ436 system. GJ436 is a well-studied M dwarf star with a known close-in planet, GJ436\,b orbiting at 0.028au \citep{butler_2004}. GJ436\,b is a warm mini-Neptune \cite[with a calculated effective temperature of $T=880$K,][]{rodgers-lee2023} and is an interesting exoplanet because it has a cometary-like outflow detected via a Ly-$\alpha$ transit with the Hubble Space Telescope \citep{ehrenreich_2015}. While these hydrogen-dominated gas giant atmospheres are less likely to be hospitable to any life-as-we-know-it in comparison to an Earth-like atmosphere, they do have an advantage: gas giant atmospheres constitute the majority of JWST and Ariel exoplanet targets. Their low mean molecular weight results in a large scale height, making them easier to observe with transmission spectroscopy than an Earth-like atmosphere with a higher mean molecular weight. 
 
 In order to model the transport of energetic particles in the GJ436 system a stellar wind model and a model for the exoplanetary atmosphere are required. In \citet{rodgers-lee2023} we use a stellar wind model from \citet{mesquita_2021}. The stellar wind properties necessary to model Galactic cosmic ray transport from the ISM into an astrosphere are the stellar wind velocity, magnetic field strength and mass loss rate. The level of turbulence present in the magnetic field, rather than the field strength itself, dictates the energetic particle transport through the magnetised stellar wind. The turbulence properties for winds other than the solar wind are unknown and we simply adopt similar properties as for the solar wind. See \citet{rodgers-lee2023} for more details.

\subsection{Energetic particle spectra at the top of the exoplanet atmosphere}
The spectrum for Galactic cosmic rays in the local ISM (known as the Local Interstellar Spectrum), outside of the solar system, has been measured by {\it Voyager 1 and 2} \citep[e.g.][]{cummings_2016}. The Galactic cosmic ray spectrum is then suppressed by the solar wind, mainly at cosmic ray energies below $\sim$30 GeV so that a somewhat reduced spectrum is detected at Earth with instruments such as {\it PAMELA} \citep[e.g.][]{potgieter_2017}. Thus, we calibrated our 1D energetic particle transport model \citep{rodgers-lee2020} using the {\it Voyager} and {\it PAMELA} detections. The LIS is used as the input spectrum outside the astrosphere of another stellar system. The energetic particle transport model from \citet{rodgers-lee2020}, based on the diffusive-advection transport equation from \citet{parker_1965}, is used in combination with the stellar wind properties \citep[Case `A' presented in][]{mesquita_2021} to calculate the Galactic cosmic ray fluxes at different orbital distances within the GJ436\,b system. These fluxes at the different orbital distances are the values we adopt as the values at the top of the hypothetical exoplanet atmospheres.

The Galactic cosmic ray fluxes are suppressed as they travel through the magnetised stellar wind because they are charged particles. The solar wind, and therefore we assume stellar winds from other low-mass stars, are turbulent. This means that Galactic cosmic rays diffuse into the solar, or stellar, system as they are deflected and bounce off perturbations in the magnetic field rather than travelling ballistically. The stellar wind is expanding, relating to its velocity, which advects the Galactic cosmic rays out of the solar system. The Galactic cosmic rays suffer adiabatic losses due to the divergence of the stellar wind's velocity field, which also can be thought of as advection in momentum space. The overall level of suppression that the Galactic cosmic rays experience in any given stellar system is a balance between these diffusive and advective processes.

\begin{figure}[t]
  \centerline{\vbox to 6pc{\hbox to 10pc{}}}
  \includegraphics[scale=.45]{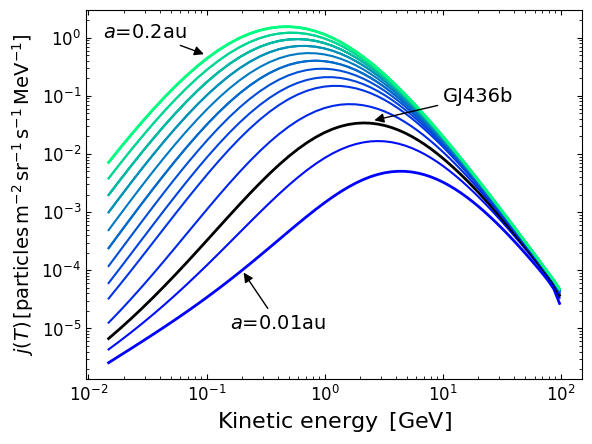}
  \includegraphics[scale=.45]{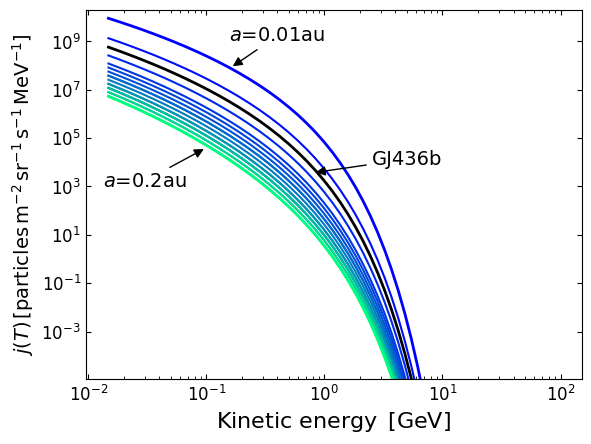}
  \caption{Differential intensities of \textit{(left)} Galactic cosmic rays and \textit{(right)} stellar energetic particles as a function of cosmic ray kinetic energy at orbital distances between 0.01 - 0.2\,au for the GJ436 system \citep[from][]{rodgers-lee2023}.}
  \label{fig:cr-fluxes}
\end{figure}

For GJ436b, the resulting Galactic cosmic ray differential intensities at orbital distances between 0.01 - 0.2\,au are shown in the left panel of Fig.\,\ref{fig:cr-fluxes}. The Galactic cosmic ray fluxes increase with increasing orbital distance. The differential intensity at the orbital distance of GJ436\,b, shown by the black line, is lower than the differential intensities observed at Earth.

The right panel of Fig.\,\ref{fig:cr-fluxes} shows the stellar energetic particle differential intensities for the same orbital distances. Again, the black line represents the differential intensities at the orbital distance of GJ436\,b. It is important to note that the scale on the left and right panels is significantly different. An interesting feature of these spectra is that the Galactic cosmic ray spectra peak at $\sim$GeV energies. In comparison for the stellar energetic particles, the intensities are much higher generally but they begin to decrease rapidly at $\sim$Gev energies. The stellar energetic particle spectra are produced based on 3 assumptions. The first is that the acceleration mechanism produces a power-law index of -2 for the low-energy part of the spectrum (indicative of diffusive shock acceleration). There is a maximum momentum that the star can accelerate stellar energetic particles to, beyond this momentum the differential intensities decrease exponentially. The final assumption is that the stellar energetic particles are constant in time.

 The exoplanet atmosphere properties (e.g. $T, p$) in \citet{rodgers-lee2023} were obtained using the HELIOS model \citep{malik_2017,malik_2019}. The temperature-pressure profiles are shown in Fig.\,3 of \citet{rodgers-lee2023}. The density is the most important atmospheric property to model energetic particle transport in exoplanet atmospheres which was calculated using the ideal gas law. In \cite{rodgers-lee2023}, for the energetic particle transport we assume a hydrogen-dominated atmosphere.
 
 \section{Analysis and results}
\begin{figure}[t]
\centering
  \centerline{\vbox to 6pc{\hbox to 50pc{50}}}
  \includegraphics[scale=.45]{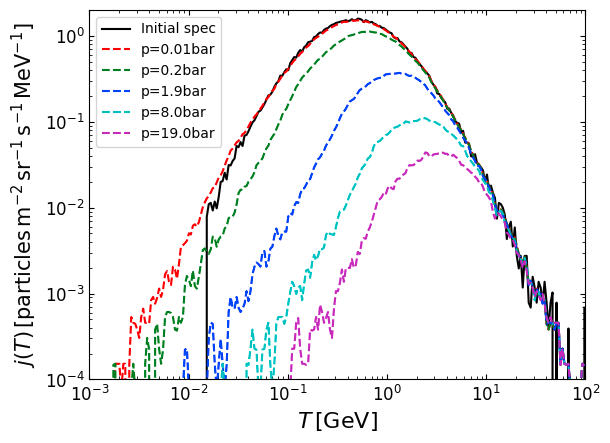}
  \caption{Differential intensities of Galactic cosmic rays as a function of cosmic ray kinetic energy at different heights in an exoplanet atmosphere for a GJ436\,b-like exoplanet at $a=0.2$\,au in the GJ436 system \citep[from][]{rodgers-lee2023}.}
  \label{fig:heights}
\end{figure}

Here, I show our results for the penetration of cosmic rays in a GJ436\,b-like atmosphere. Each of the spectra shown in Fig.\,\ref{fig:cr-fluxes} is propagated down through an exoplanet atmosphere using a Monte Carlo cosmic ray transport code \citep{rimmer_2013}. For example, the solid black line in Fig.\,\ref{fig:heights} shows the initial Galactic cosmic ray spectrum at the top of the atmosphere for $a=0.2$au. The dashed coloured lines show the subsequent spectra for increasing pressures in the atmosphere as the Galactic cosmic rays lose energy by interacting with the atmosphere. The resulting spectra for the Galactic cosmic rays and the stellar energetic particles is then used to calculate the ionisation rate of molecular hydrogen at each height in the atmosphere.

\subsection{Ionisation rate of molecular hydrogen}
The ionisation rate of molecular hydrogen, $\zeta$, due to energetic particles gives an indication of how important, and where, they might be at driving chemistry in the exoplanet atmosphere. The $\mathrm{H_2}$ ionisation rate is calculated from the differential intensities and the $\mathrm{H_2}$ ionisation cross-section \citep[Eq.4 from][]{rodgers-lee2023}.

\begin{figure}[t]
  \centerline{\vbox to 6pc{\hbox to 10pc{}}}
  \includegraphics[scale=.45]{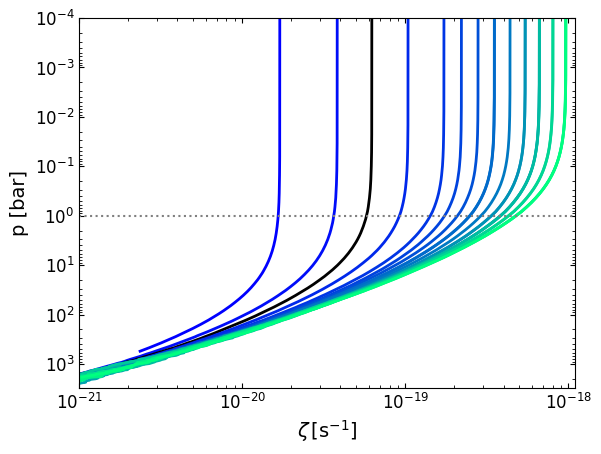}
  \includegraphics[scale=.45]{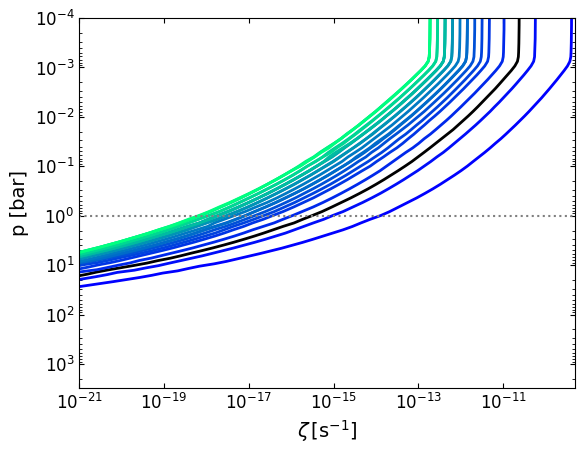}
  \caption{The ionisation rate of molecular hydrogen due to \textit{(left)} Galactic cosmic rays and \textit{(right)} stellar energetic particles as a function of atmospheric pressure for a GJ436\,b-like planet at orbital distances between 0.01 - 0.2\,au for the GJ436 system \citep[adapted from][]{rodgers-lee2023}.}
  \label{fig:zeta}
\end{figure}

Fig.\,\ref{fig:zeta} shows the ionisation rate of molecular hydrogen, $\zeta$, as a function of pressure for a GJ436\,b-like planet for $a=0.01-0.2$\,au \citep[adapted from][]{rodgers-lee2023}. The left and right panels plot $\zeta$ as a result of ionisation due to Galactic cosmic rays and stellar energetic particles, respectively. The black solid line represents $\zeta$ in an exoplanet atmosphere at 0.028\,au, the orbital distance of GJ436\,b. The dotted grey line denotes $P=1$\,bar for reference. For comparison, the LIS results in $\zeta\sim 10^{-17}\mathrm{s^{-1}}$.

Additionally, for both Galactic cosmic rays and stellar energetic particles, the value of $\zeta$ remains constant with increasing pressure up to a certain point. In this region of the atmosphere the energetic particles are not losing a significant amount of their energy. For Galactic cosmic rays, $\zeta$ begins to decrease between $p\sim 10^{1} - 10^{-1}$bar for $a=0.01 - 0.2$au, respectively. The stellar energetic particles are absorbed at much higher heights in the atmosphere, corresponding to $p\sim10^{-2}$\,bar. Thus, stellar energetic particles dominate as a source of ionisation over Galactic cosmic rays in the upper atmosphere. At high pressures, $p\gtrsim 1$bar, although overall the ionisation rate is low Galactic cosmic rays dominate instead. Transmission spectroscopy with JWST and Ariel will probe between $10^{-4}-10^{-1}$bar in hydrogen dominated atmospheres \citep{welbanks2019}. Thus, transmission spectroscopy with JWST and Ariel is more likely to be sensitive to chemistry driven by stellar energetic particles for the atmospheres and orbital distances studied here.

\section{Discussion and Summary}
I presented our results of the ionising effect of stellar energetic particle and Galactic cosmic rays on a hydrogen-dominated exoplanet atmosphere for $a=0.01-0.2$au in the M dwarf system, GJ436. This includes the orbital distance of the known mini-Neptune, GJ436\,b. I discussed how the stellar wind properties, along with the composition and density of an exoplanet atmosphere determine the energetic particle fluxes in the exoplanet atmosphere. Stellar energetic particles lead to high ionisation rates in a GJ436\,b-like atmosphere at the orbital distances discussed here, namely between 0.01-0.2\,au. The Galactic cosmic ray ionisation rates for a hypothetical exoplanet at these orbital distances were generally much lower but dominate over stellar energetic particles deep in the exoplanet atmosphere.

The results presented here did not include the effect of a planetary magnetic field. A planetary magnetic field would shield the atmosphere by deflecting both stellar energetic particles and Galactic cosmic rays, up to a certain energy, towards the poles \citep[see e.g.][]{herbst_2019}. Investigating the effect of a planetary magnetic is of great interest. As mentioned above, the stellar energetic particle fluxes are assumed to be constant here. In reality, stellar flares and CMEs are unlikely to produce constant fluxes of associated stellar energetic particles that impact an exoplanet between $a=0.01-0.2$au for this system, given that it is not an active star. It is not clear what impact this would have on chemistry in the atmosphere at this point. Chemical modelling of an exoplanet atmosphere including energetic particle ionisation are needed to address this time dependent effect.

Using chemical models, \citet{helling_2019} identified $\mathrm{H_3^+}$ and $\mathrm{H_3O^+}$ as ions indicative of Galactic cosmic ray ionisation in a brown dwarf/gas giant atmosphere. The important pathway for creating these ions is:
\begin{eqnarray}
\mathrm{ H_2 + CR }\rightarrow \mathrm{ H_2^+ + e^- + CR}\\
\mathrm{H_2^+ +H_2 }\rightarrow \mathrm{H_3^+ + H} \\
\mathrm{H_3^+ + H_2O} \rightarrow \mathrm{ H_3O^+ + H_2}
\end{eqnarray}
where `CR' is short for cosmic ray. Stellar energetic particles will produce similar effects \citep[e.g.][]{barth_2021}. Note, the presence of water is important. Scaling the abundances of $\mathrm{H_3^+}$ and $\mathrm{H_3O^+}$ from \citet{helling_2019}, \citet{bourgalais_2020} produced synthetic JWST and Ariel transmission spectra for a GJ1214\,b-like atmosphere, showing that absorption features from $\mathrm{H_3^+}$ and $\mathrm{H_3O^+}$ in the near-infrared would be strong enough to be observed with Ariel and NIRSpec on JWST (see their Fig.\,6). This remains to date the only synthetic JWST and Ariel transmission spectrum for a hydrogen-dominated atmosphere that includes $\mathrm{H_3^+}$ and $\mathrm{H_3O^+}$, formed as a result of energetic particle ionisation. 

The stellar energetic particle and Galactic cosmic ray spectra shown here can be used in the future with chemical and radiative transfer models to produce a suite of synthetic transmission spectra, building on the results presented in \citet{bourgalais_2020}. This would identify the energetic particle ionisation rate needed to result in sufficient abundances of $\mathrm{ H_3^+}$ and $\mathrm{ H_3O^+}$ to be detected with transmission spectroscopy. In turn this would help us identify the types of exoplanets to target, in terms of orbital distance and stellar activity, to detect these features. Alongside this more direct connection with observations, more sophisticated 2D/3D energetic particle transport models can be employed in the future \citep[see e.g.][]{engelbrecht_2024}. 

\section{Acknowledgements}
I would like to thank the Scientific Organising Committee for the invitation to take part in the IAU symposium 388. DRL would like to acknowledge that this publication has emanated from research conducted with the financial support of Science Foundation Ireland under Grant number 21/PATH-S/9339.

\end{document}